\documentstyle[aps,prb,multicol,epsf]{revtex}

\begin{document}

\title{
Metal-to-Insulator Transition, Spin Gap Generation, and Charge Ordering
in Geometrically Frustrated Electron Systems
}

\author{Satoshi Fujimoto}
\address{
Department of Physics,
Kyoto University, Kyoto 606-8502, Japan
}

\date{\today}
\maketitle
\begin{abstract}
We investigate a (semi-)metal to insulator transition (MIT) realized in
geometrically frustrated electron systems on the basis of
the Hubbard model on a three-dimensional pyrochlore lattice 
and a two-dimensional checkerboard lattice.
Using the renormalization group method 
and mean field analysis,
we show that in the half-filling case, 
MIT occurs as a result of the interplay between geometrical frustration
and electron correlation.
In the insulating phase, which has a spin gap, the spin rotational
symmetry is not broken, while charge ordering exists.
The charge ordered state is stabilized so as to relax the geometrical
frustration in the spin degrees of freedom.
We also discuss the distortion of the lattice structure caused by
the charge ordering.
The results are successfully applied to the description of 
the MIT observed in the pyrochlore system 
${\rm Tl_2Ru_2O_7}$.
\end{abstract}

\pacs{PACS numbers: }

\begin{multicols}{2}

\section{Introduction}

Geometrical frustration in both localized and
itinerant electron systems is an important ingredient giving rise to
a rich variety of condensed phases.
\cite{rami,pyroex,gau,har,maeno1,fuka,ue,col}
In localized spin systems, magnetic frustration suppresses
a tendency toward a conventional long-range order, and may
stabilize some exotic state such as a spin liquid
or a valence bond crystal, as has been extensively explored
both experimentally\cite{rami,pyroex,gau,har,maeno1,fuka,ue,col} 
and theoretically.
\cite{py,re,can,ber,iso,ya,ko,ts,check,lieb,sta,auer,fou,kag,chan,lech,tch}
Also, for itinerant systems, it has been argued that
geometrical frustration may cause some novel phenomena such as  
heavy fermion state,\cite{li,fuji,iso2,fuji2,ts2,ya2,cole} 
anomalous Hall effect 
induced by the chiral order,\cite{naga,kawa} and so forth.
Among them, metal-to-insulator transition (MIT) in geometrically
frustrated systems is an intriguing unsettled issue. 
It has been found experimentally that 
the pyrochlore oxides, ${\rm Tl_2Ru_2O_7}$ and ${\rm Cd_2Os_2O_7}$,
and the spinel compounds, ${\rm CuIr_2S_4}$ and ${\rm MgTi_2O_4}$,
exhibit MIT without
magnetic long-range order at finite critical
temperatures.\cite{take,mand,furu,isobe}  
Since such systems possess the fully-frustrated lattice structure, 
referred to as a network of corner-sharing tetrahedra (that is, a pyrochlore
lattice), the magnetic properties of 
the insulating phase are not yet understood.
Moreover the mechanism of the MITs observed in these systems 
is still an open problem.
In contrast to the localized spin systems, 
the presence of charge degrees of
freedom provides a route for the relaxation of magnetic frustration.
However, when electron correlation is sufficiently strong,  
the magnetic frustration may still affect the low-energy properties 
significantly.
Thus, it is expected that
geometrical frustration plays an important role in the MITs.
From this point of view, in the present paper, 
we study the interplay between electron correlation
and geometrical frustration in the Hubbard model 
on a three-dimensional (3D) pyrochlore lattice
and on a two-dimensional (2D) checkerboard lattice,
the so-called 2D pyrochlore (FIG.1).
Although real pyrochlore oxides and spinel compounds have electronic structure
composed of $t_{2g}$ orbitals, the present study on these simpler
single-band models
may provide important insight into the role of geometrical frustration
in MIT. 
Furthermore, ${\rm Tl_2Ru_2O_7}$ has, apart from the $t_{2g}$ band, 
a nearly half-filled Tl $s$ band, 
whose important features are described 
by the 3D pyrochlore Hubbard model.\cite{ogu}
We believe that this model may provide a useful understanding of 
the MIT undergone in this material.


\begin{figure}[h]
\centerline{\epsfxsize=7.5cm \epsfbox{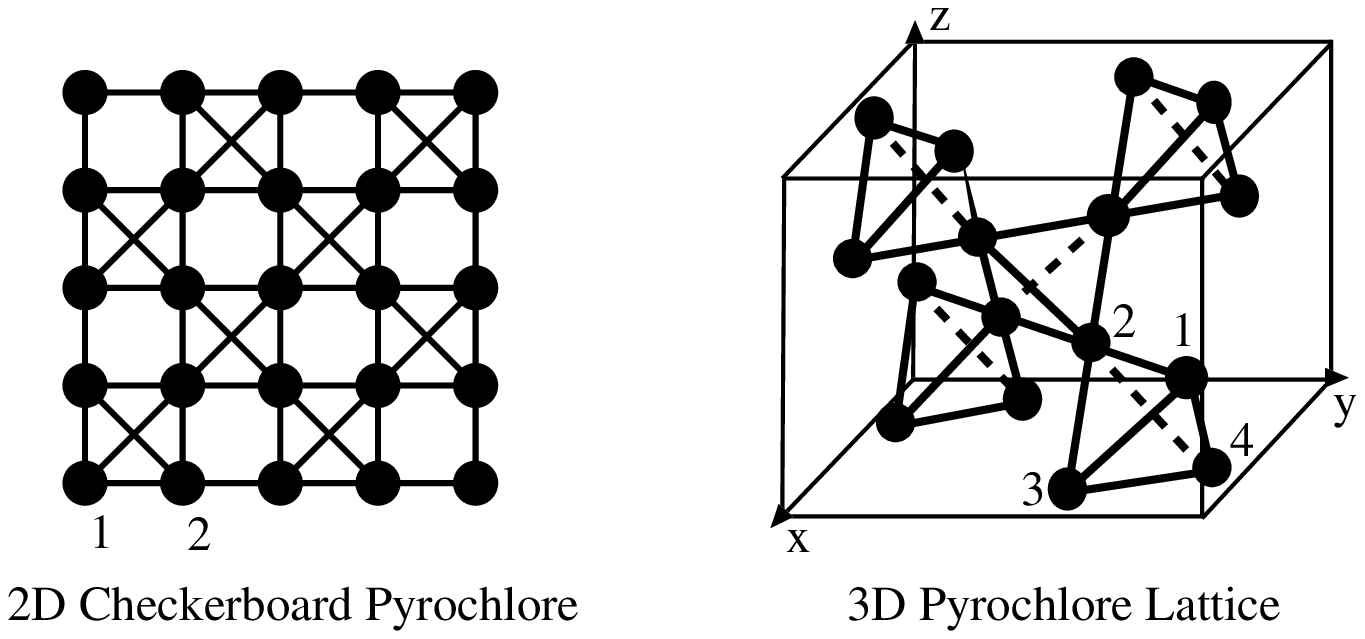}}
{FIG. 1. 2D and 3D pyrochlore lattices.}
\end{figure}

The non-interacting energy bands of these two Hubbard models 
have a common interesting feature: 
They consist of a flat band (or two degenerate flat bands) 
on the upper band edge and
a dispersive band that is tangent to the flat band 
(or flat bands) at the $\Gamma$ point.\cite{mielke}
Itinerant ferromagnetism 
in the case that the flat bands are partially
occupied has been extensively studied 
by several authors so far.\cite{mielke,ferro}
This particular band structure is due to
the geometrical property of {\it line graph} in which  
the above 2D and 3D pyrochlore lattices are classified.
According to a theorem of the graph theory,\cite{graph} 
a tight-binding model on a line graph generated from
a bipartite graph with $V$ vertices and $E$ edges has 
flat bands with the degeneracy $D_f=\lim_{N\rightarrow\infty}(E-V+1)/N$
lying on the band edge of dispersive bands.
Here $N$ is the total number of unit cells on the lattice.
It should be noted that 
the existence of the flat bands is analogous to the macroscopically large
degeneracy of the ground state of classical spin systems 
on line graphs.\cite{py}
Since line graphs consist of a network of complete graphs which have
a frustrated structure, the presence of the flat bands is a result of
geometrical frustration inherent in the lattice structure.
We would like to stress that, to study effects of geometrical frustration 
in correlated electron systems,
we should distinguish two classes of geometrically frustrated systems:
one is the class of line graph, and the other is not, and does not
possess flat bands.  
For example, a triangular lattice is classified in the latter class.
In this paper, we are concerned with the former class, and
show that, in the half-filling case, 
the geometrical frustration in the above models drives the system 
into an insulating state with a spin gap as well as a charge gap.
Also, in the insulating state, the spin rotational symmetry is not broken,
while charge ordering exists.
The results of this paper have been partially reported in Ref.43
before. 

The paper is organized as follows.
In Sec. II, we introduce the model and present briefly 
some basic results derived by
perturbative expansion in terms of electron-electron interaction.
In Sec. III, the renormalization group method applied to
the pyrochlore Hubbard models is developed.
The results for the 2D and 3D cases are given in Sec.IV and V, respectively.
In particular, we discuss the (semi-)metal-to-insulator transition caused by
geometrical frustration, and the properties of the insulating state
in which both spin and charge excitations have a gap.
We confirm our results with the use of the mean field analysis.
The effect of the coupling with lattice degrees of freedom
in the presence of the charge ordering is also discussed.
Summary is given in Sec. VI.

\section{Model Hamiltonians and Some Perturbative Results}

Our system is described by the single-band Hubbard model 
defined on the 2D checkerboard lattice
or the 3D pyrochlore lattice shown in FIG.1.
Diagonalizing the kinetic term, we write the Hamiltonian as
\begin{eqnarray}
&&H=\sum_{\mu=1}^{m}\sum_{k\sigma}E_{k\mu}a^{\dagger}_{k\mu\sigma}
a_{k\mu\sigma} \nonumber \\
&&\qquad +\frac{U}{N}\sum_{k,k',q}\sum_{\alpha\beta\gamma\delta}
\Gamma^0_{\alpha\beta\gamma\delta}(k-q,k'+q;k',k) \nonumber \\
&&\qquad\times a^{\dagger}_{k-q\alpha\uparrow}
a^{\dagger}_{k'+q\beta\downarrow}
a_{k'\gamma\downarrow}a_{k\delta\uparrow}, \label{hamil} \\
&&\Gamma^0_{\alpha\beta\gamma\delta}(k_1,k_2;k_3,k_4)= 
\sum_{\nu=1}^m 
s_{\nu\alpha}(\mbox{\boldmath $k_1$})
s_{\nu\beta}(\mbox{\boldmath $k_2$})
s_{\nu\gamma}(\mbox{\boldmath $k_3$})
s_{\nu\delta}(\mbox{\boldmath $k_4$}), \nonumber 
\end{eqnarray}
where $m=2$ in the 2D case, and 
$m=4$ in the 3D case. $a_{k\mu\sigma}$ ($a^{\dagger}_{k\mu\sigma}$)
is the annihilation (creation) operator for electron with momentum $k$
and spin $\sigma$ in the $\mu$-band.

In the 2D case, 
\begin{eqnarray}
E_{k1}&=&2, \\
E_{k2}&=&4\cos k_x \cos k_y-2,
\end{eqnarray}
\begin{eqnarray} 
&&s_{11}(\mbox{\boldmath $k$})=s_{22}(\mbox{\boldmath $k$})
=\frac{\sin[(k_x+k_y)/2]}{\sqrt{1-\cos k_x\cos k_y}}, \\
&&s_{21}(\mbox{\boldmath $k$})=-s_{12}(\mbox{\boldmath $k$})
=\frac{\sin[(k_x-k_y)/2]}{\sqrt{1-\cos k_x\cos k_y}}.
\end{eqnarray}

In the 3D case, 
\begin{eqnarray}
E_{k1}&=&E_{k2}=2, \\
E_{k3}&=&-2+2\sqrt{1+t_k}, \\
E_{k4}&=&-2-2\sqrt{1+t_k},  \\
t_k&=&\cos(2k_x)\cos(2k_y)+\cos(2k_y)\cos(2k_z) \nonumber \\
&&+\cos(2k_z)\cos(2k_x).
\end{eqnarray}
Using the abbreviation, $s(x\pm y)\equiv \sin(k_x\pm k_y)$ etc., we write down 
$s_{\mu 1},s_{\mu 2}$(k) in the 3D case,
\begin{eqnarray}
&&(s_{11}(k),s_{21}(k),s_{31}(k),s_{41}(k)) \nonumber \\
&&=(s(x+z),s(y-z),-s(x+y),0)/n_{+},
\end{eqnarray}
where $n_{\pm}=\sqrt{s(x\pm z)^2+s(y\mp z)^2+s(x+y)^2}$, and,
\begin{eqnarray}
&&(s_{12}(k),s_{22}(k),s_{32}(k),s_{42}(k)) \nonumber \\
&&=(s(x+z)s(x-z)s(y-z) \nonumber \\
&&-s(y+z)(s(y-z)^2+s(x+y)^2), \nonumber \\
&& s(x+z)s(y+z)s(y-z)  \nonumber \\
&&-s(x-z)(s(x+z)^2+s(x+y)^2), \nonumber \\
&& -s(x+y)(s(x+z)s(y+z)+s(x-z)s(y-z)), \nonumber \\
&& s(x+y)n_{+}^2)/n_2,
\end{eqnarray}
for $k_x+k_y\neq 0$, where $n_2=n_{+}[n_{+}^2n_{-}^2-(s(x+z)s(y+z)
+s(x-z)s(y-z))^2]^{\frac{1}{2}}$, and,
\begin{eqnarray}
&&(s_{12}(k),s_{22}(k),s_{32}(k),s_{42}(k)) \nonumber \\
&&=(-s(2x),-s(2x),2s(x-z),2s(x+z)) \nonumber \\
&\times& 1/\sqrt{2s(2x)^2+4s(x-z)^2+4s(x+z)^2}
\end{eqnarray}
for $k_x+k_y=0$.
The expressions of $s_{\mu 3}(k)$ and $s_{\mu 4}(k)$ are very complicated.
However in the following, we need only $s_{\mu 3}(k)$ for small $k$, 
which is given by,
\begin{eqnarray}
&&(s_{13}(k),s_{23}(k),s_{33}(k),s_{43}(k)) \nonumber \\
&&=(-k_x-k_y+k_z,k_x+k_y+k_z, \nonumber \\
&& -k_x+k_y-k_z,k_x-k_y-k_z)
/2\vert k\vert.
\end{eqnarray}
The annihilation operator of electrons at the $\mu$-th site
in a unit cell is given by 
$c_{k\mu\sigma}=\sum_{\nu=1}^ms_{\mu\nu}(\mbox{\boldmath $k$})a_{k\nu\sigma}$.

As mentioned in the introduction, these systems have the flat band(s),
$E_{k1}$ for the 2D case, and $E_{k1}$, $E_{k2}$ for the 3D case.
In the half-filling case, $n=1$, on which we concentrate henceforth,
the flat band(s) is empty, while
the dispersive band(s) below the flat band(s) is fully occupied. 
In the non-interacting half-filling case,
the system is in a semi-metal state,
since the Fermi velocity is vanishing, though there is no excitation gap. 
In the following, we study how this state is affected by electron correlation.

As shown in Ref.27, at the half-filling,
the perturbative calculation in $U$ for the Hamiltonian (\ref{hamil}) 
suffers from divergences of the single-particle self-energy, 
due to the presence of the flat band(s).
In the 2D case, the perturbative expansion of the self-energy gives,
\begin{eqnarray}
{\rm Re}\Sigma^{\rm 2D}(\varepsilon)\sim {\rm const.}
+cU^2\log(8t/\varepsilon)+\cdot\cdot\cdot.
\end{eqnarray} 
In the 3D case, we have,
\begin{eqnarray}
{\rm Re}\Sigma^{\rm 3D}(\varepsilon)&\sim& 
{\rm const.}+c_2U^2\sqrt{|\varepsilon|}
+c_3U^4/\sqrt{|\varepsilon|} \nonumber \\
&&+c_4 U^4/|\varepsilon|^{3/2}+\cdot\cdot\cdot.
\end{eqnarray}
These singular behaviors imply that some instability may be induced by
electron correlation.
To pursue this possibility, we will carry out the resummation of 
divergent terms using the renormalization group method in the following
sections.

\section{Renormalization Group Equations}

To treat the infrared divergences appeared in perturbative expansion 
in a controlled manner,
we exploit the renormalization group (RG) method.
In previous application of the RG method to 
electron systems,\cite{shan,metz,kop,zan,hone} 
a momentum cutoff that separates 
the neighborhood of the Fermi surface from the higher momentum part
is introduced.
However, in the presence of the flat band(s), this procedure is not applicable.
To circumvent this problem, 
we introduce the infrared energy cutoff $\Lambda$ in the following manner:
\begin{eqnarray}
\psi_{\mu\sigma}(k,\varepsilon_n)
&=&\psi^{>}_{\mu\sigma}(k,\varepsilon_n)
\Theta(|\varepsilon_n|-\Lambda) \nonumber \\
&&+\psi^{<}_{\mu\sigma}(k,\varepsilon_n)\Theta(\Lambda-|\varepsilon_n|).
\end{eqnarray}
Here, $\psi_{\mu\sigma}(k,\varepsilon)$ 
is the Grassmann field corresponding to $a_{k\mu\sigma}$.

Using a standard method,\cite{frg,wet,morris,salm} we obtain
the RG equations of the single-particle
self-energy for electrons in the $\mu$ and $\nu$ bands, 
$\Sigma_{\mu\nu}(k)$, and the 4-point vertex functions
for electrons in the $\alpha$, $\beta$, $\gamma$, and $\delta$ bands,
$\Gamma_{\alpha\beta\gamma\delta}(k_1,k_2;k_3,k_4)$:
\begin{eqnarray}
&&\frac{\partial \Sigma^{\Lambda}_{\mu\nu}(k)}{\partial \Lambda}=
-\sum_{k'}\delta(|\varepsilon_n'|-\Lambda)
G^{\Lambda}_{\alpha\beta}(k') \nonumber \\
&&\qquad\qquad\qquad
\times\Gamma^{(4)\Lambda}_{\mu\beta\nu\alpha}(k,k';k,k'), \label{selfre}\\
&&\frac{\partial \Gamma^{(4)\Lambda}_{\alpha\beta\gamma\delta}
(k_1,k_2;k_3,k_4)}{\partial \Lambda}=\sum_{k,k'}\sum_{\mu\nu\lambda\kappa}
[\Theta(|\varepsilon_n|-\Lambda)\delta(|\varepsilon_n'|-\Lambda)
\nonumber \\
&&+\Theta(|\varepsilon_n'|-\Lambda)\delta(|\varepsilon_n|-\Lambda)]
G^{\Lambda}_{\mu\lambda}(k)G^{\Lambda}_{\nu\kappa}(k')
\nonumber \\
&&\times[\frac{1}{2}\Gamma^{(4)\Lambda}_{\alpha\beta\mu\nu}(k_1,k_2;k,k')
\Gamma^{(4)\Lambda}_{\lambda\kappa\gamma\delta}(k,k';k_3,k_4)
\delta_{k_1+k_2,k+k'}  \nonumber \\
&&-\Gamma^{(4)\Lambda}_{\alpha\mu\gamma\nu}(k_1,k;k_3,k')
\Gamma^{(4)\Lambda}_{\kappa\beta\lambda\delta}(k',k_2;k,k_4)
\delta_{k_1+k,k_3+k'}  \nonumber \\
&&+\Gamma^{(4)\Lambda}_{\alpha\mu\delta\nu}(k_1,k;k_4,k')
\Gamma^{(4)\Lambda}_{\kappa\beta\lambda\gamma}(k',k_2;k,k_3)
\delta_{k_1+k,k_4+k'}] \nonumber \\
&&+\sum_{k}\delta(|\varepsilon_n|-\Lambda)
G^{\Lambda}_{\mu\nu}(k)\Gamma^{(6)\Lambda}_{\alpha\beta\mu\nu\gamma\delta}
(k_1,k_2,k,k,k_3,k_4). \label{vert}
\end{eqnarray}
Here $G^{\Lambda}_{\mu\nu}(k)=[({\rm i}\varepsilon_n-E_{k\mu})
\delta_{\mu\nu}-\Sigma_{\mu\nu}^{\Lambda}(k)]^{-1}$ and 
$k=({\rm i}\varepsilon_n,\mbox{\boldmath $k$})$ and so forth. 
$\Gamma^{(6)\Lambda}$ is the 6-point vertex.
The first, second, and third terms of the right-hand side
of (\ref{vert}) correspond to BCS, ZS, and ZS' processes, respectively,
of which the diagrammatic expressions are shown in FIG.2(a).

\begin{figure}[h]
\centerline{\epsfxsize=7.5cm \epsfbox{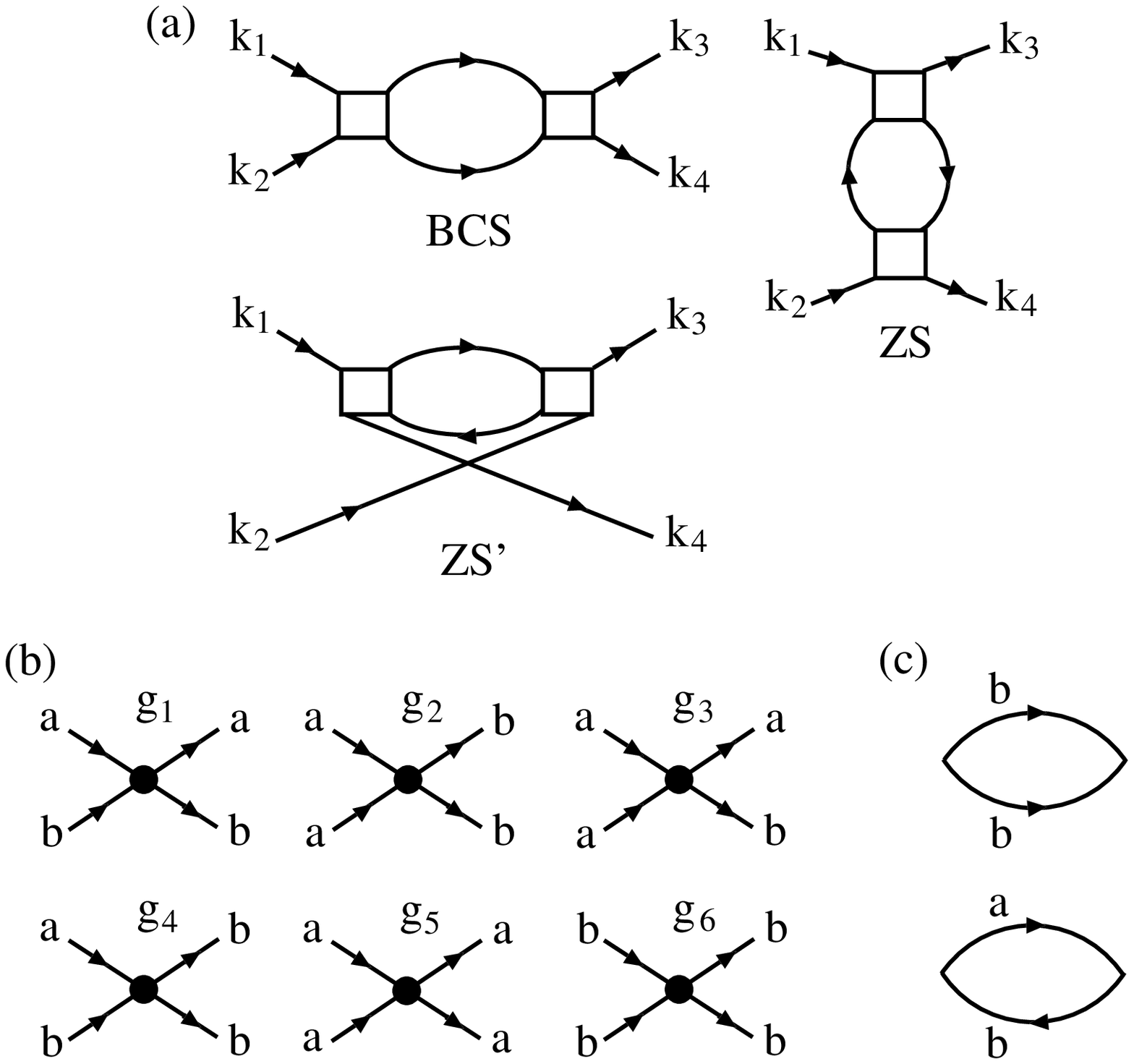}}
{FIG. 2(a) BCS, ZS, and ZS' processes.
(b) The six species of 4-point vertices. Here
``$a$'' and ``$b$'' indicate 
the dispersive band and the flat band, respectively. 
(c) The leading singular bubble diagrams.}
\end{figure}

In the following, we investigate the RG flow of the 4-point
vertex functions up to one-loop order, and drop the 6-point vertex 
$\Gamma^{(6)\Lambda}$ and the self-energy $\Sigma^{\Lambda}(k)$
in the right-hand side of (\ref{vert}).
In our systems, there are six species of 4-point vertices, 
as shown in FIG.2(b), apart from the spin degrees
of freedom and the two-fold degeneracy of the flat bands in the 3D case. 
We assume that the momentum dependences of the 4-point vertex functions are
given mainly by $\Gamma^0(k_1,k_2;k_3,k_4)$ 
in the renormalization processes. 
This is made explicitly by replacing $\Gamma_{abab}(k_1,k_2;k_3,k_4)$ with
$g_1\Gamma^0_{abab}(k_1,k_2;k_3,k_4)$, $\Gamma_{bbba}(k_1,k_2;k_3,k_4)$
with $g_4\Gamma^0_{bbba}(k_1,k_2;k_3,k_4)$, etc.
This approximation is fairly good, because 
in the vicinity of the $\Gamma$ point,
where the most important scattering processes occur,
the band structure is almost isotropic.
The spin degrees of freedom of the running couplings $g_{i}$ ($i=1\sim 6$)
is incorporated by decomposing $g_{i}$
into the spin singlet part and the spin triplet part,
\begin{eqnarray}
g_i(\sigma_1,\sigma_2;\sigma_3,\sigma_4)&=&\frac{g_{i s}}{2}
(\delta_{\sigma_1\sigma_3}\delta_{\sigma_2\sigma_4}
-\delta_{\sigma_1\sigma_4}\delta_{\sigma_2\sigma_3}) \nonumber \\
&&+\frac{g_{i t}}{2}(\delta_{\sigma_1\sigma_3}\delta_{\sigma_2\sigma_4}
+\delta_{\sigma_1\sigma_4}\delta_{\sigma_2\sigma_3}).
\end{eqnarray}
It is also convenient in the following to decompose $g_i$ 
into the charge part and the spin part,
\begin{eqnarray}
g_i(\sigma_1,\sigma_2;\sigma_3,\sigma_4)&=&\frac{3g_{i t}-g_{i s}}{4}
\delta_{\sigma_1\sigma_4}\delta_{\sigma_2\sigma_3} \nonumber \\
&&+\frac{g_{i s}+g_{i t}}{4}\mbox{\boldmath $\sigma$}_{\sigma_1\sigma_4}
\cdot\mbox{\boldmath $\sigma$}_{\sigma_2\sigma_3}.
\end{eqnarray}

Because, at the half-filling, 
the flat bands are empty and the dispersive band
is fully occupied, the particle-particle processes between the flat bands 
and the particle-hole processes
between the flat bands and the dispersive band give the leading
singular contributions (see FIG.2(c)). 
We take into account these contributions in the derivation
of the RG equations.
It is straightforward to show that 
in the vicinity of the $\Gamma$ point the following relations hold:  
\begin{eqnarray}
\sum_{\rm BCS}\Gamma^0_{aabb}\Gamma^0_{bbab}&\approx& 0,\qquad
\sum_{\rm ZS}\Gamma^0_{aabb}\Gamma^0_{baaa}\approx 0,  \nonumber \\
\sum_{\rm ZS'}\Gamma^0_{aabb}\Gamma^0_{baaa}&\approx& 0, \qquad
\sum_{\rm ZS}\Gamma^0_{abab}\Gamma^0_{aaab}\approx 0,  \nonumber \\
\sum_{\rm ZS'}\Gamma^0_{abab}\Gamma^0_{aaab}&\approx& 0. \qquad
\sum_{\rm ZS}\Gamma^0_{aaab}\Gamma^0_{bbab}\approx 0,  \nonumber \\
\sum_{\rm ZS}\Gamma^0_{aabb}\Gamma^0_{bbba}&\approx& 0, \qquad
\sum_{\rm ZS'}\Gamma^0_{aabb}\Gamma^0_{bbba}\approx 0, \nonumber \\ 
\sum_{\rm ZS}\Gamma^0_{abbb}\Gamma^0_{bbab}&\approx& 0,  \qquad
\sum_{\rm ZS'}\Gamma^0_{abbb}\Gamma^0_{bbab}\approx 0.  
\end{eqnarray}
Here $\sum_{\rm BCS}$, $\sum_{\rm ZS}$, and $\sum_{\rm ZS'}$ 
mean the momentum summation in the intermediate state
carried out over BCS, ZS, and ZS' processes, respectively.
Using these relations, we find that the beta function of $g_3$
is approximately vanishing,
\begin{eqnarray}
\frac{d g_{3s}}{d l}\approx \frac{d g_{3t}}{d l}\approx 0.
\end{eqnarray}
and the RG equations for the other running couplings are written,\cite{corr} 
\begin{eqnarray}
\frac{d g_{1s}}{d l}&=&
-\frac{ag_{4s}^2e^l}{\Lambda_0}+\frac{b(\Lambda_0e^{-l})^{\eta}}{4}
(g_{1s}^2+6g_{1s}g_{1t}-3g_{1t}^2), 
\label{rge1}\\
\frac{d g_{1t}}{d l}&=&
\frac{b(\Lambda_0e^{-l})^{\eta}}{4}
(g_{1s}^2-2g_{1s}g_{1t}+5g_{1t}^2), 
\label{rge2}\\
\frac{d g_{2s}}{d l}&=&
-\frac{ag_{2s}g_{6s}e^l}{\Lambda_0}+b(\Lambda_0e^{-l})^{\eta}
[g_{1s}g_{2s} \nonumber \\
&&+3(g_{1s}g_{2t}+g_{1t}g_{2s}-g_{1t}g_{2t})], 
\label{rge5}\\
\frac{d g_{2t}}{d l}&=&-\frac{ag_{2t}g_{6t}e^l}{\Lambda_0}, \label{rge6} \\
\frac{d g_{4s}}{d l}&=&
-\frac{ag_{4s}g_{6s}}{\Lambda_0}e^l
+\frac{b(\Lambda_0e^{-l})^{\eta}}{2}
(g_{1s}g_{4s}+3g_{4s}g_{1t}), \label{rge3}\\
\frac{d g_{4t}}{d l}&=&
-\frac{ag_{4t}g_{6t}}{\Lambda_0}e^l, \label{rge7} \\
\frac{d g_{5s}}{d l}&=&-\frac{ag_{2s}^2e^l}{\Lambda_0}, \label{rge8} \\
\frac{d g_{5t}}{d l}&=&-\frac{ag_{2t}^2e^l}{\Lambda_0}, \label{rge9} \\
\frac{d g_{6s}}{d l}&=&
-\frac{ag_{6s}^2}{\Lambda_0}e^l, \label{rge4} \\
\frac{d g_{6t}}{d l}&=&
-\frac{ag_{6t}^2}{\Lambda_0}e^l, \label{rge10}
\end{eqnarray}
where $\eta=(d-2)/2$, 
$l={\rm ln}(\Lambda_0/\Lambda)$, with $\Lambda_0$ the band width, 
and $d$ is the spatial dimension.
In the 2D case, 
\begin{eqnarray}
a&=&\sum_k(s_{11}^4(k)-s_{11}^2(k)s_{12}^2(k))/2, \\
b&=&\sum_k{\rm Im}[i\Lambda-E_{k2}]^{-1}/2\approx 0.0622/t,
\end{eqnarray} 
and in the 3D case, 
\begin{eqnarray}
a&=&\sum_k[(s_{11}^2(k)+s_{12}^2(k))^2  \nonumber \\
&&-(s_{11}(k)s_{21}(k)+s_{12}(k)s_{22}(k))^2]/2, \\  
b&=&\sum_k{\rm Im}[i\Lambda-E_{k3}]^{-1}/(4\sqrt{\Lambda})
\approx 0.0775/t^{3/2}.
\end{eqnarray}
In the derivation of these equations for the 3D case, 
we have used the fact that in the vicinity of the $\Gamma$ point,
the two degenerate flat bands 
do not mix with each other in scattering processes. 
Thus in this case the two-fold degeneracy just gives 
an overall factor of $2$.

Since the initial values of the running couplings in the triplet channel
are zero for our models (\ref{hamil}), the RG equations, 
(\ref{rge6}), (\ref{rge7}), (\ref{rge9}), and (\ref{rge10}), 
give $g_{2t}=g_{4t}=g_{5t}=g_{6t}=0$.
In the following sections, 
we study how the RG flows of the other running couplings 
give rise to non-trivial effects.

\section{2D Pyrochlore Hubbard Model}

\subsection{RG analysis}

We first consider the 2D case, whose theoretical treatment is simpler.
We solved the RG equations (\ref{rge1})-(\ref{rge10}) numerically
for a particular set of parameter values, and
obtained the RG flow shown in FIG.3. 
The running couplings $g_{5s}$ and $g_{6s}$, 
of which the RG flows are not shown in FIG.3,
are irrelevant in the low-energy limit. 
We found that for any small value of $U/t$, 
$g_{1t}$ flows into the strong-coupling regime. This indicates some
instability in this channel.
Although $g_{1s}$ also scales into the strong-coupling regime, 
it is sub-dominant compared to $g_{1t}$.
We also show in FIG.3 the RG flows of the couplings $3g_{1t}-g_{1s}$ 
and $g_{1s}+g_{1t}$, which are related to the charge and spin
susceptibilities, respectively.
We see that some instability appears in the charge degrees of freedom.
To elucidate the nature of this instability more precisely, 
we explore how this singularity affects the single-particle self-energy.
Although the diagonal parts of the self-energy $\Sigma_{11}$, $\Sigma_{22}$
give just a chemical potential shift up to the one loop level,
the off-diagonal self-energy $\Sigma_{12}$ changes
the electronic state drastically, as will be shown below.
Neglecting the diagonal self-energy, which are not important
in the following argument, 
we expand the RG equation 
for the off-diagonal self-energy (\ref{selfre}) 
up to the first order in $\Sigma^{\Lambda}_{12}$:
\begin{eqnarray}
&&\frac{d(\Sigma_{12\uparrow\uparrow}^{\Lambda}
+\Sigma_{12\downarrow\downarrow}^{\Lambda})}{d l}
=2b(3g_{1t}-g_{1s})
(\Sigma_{12\uparrow\uparrow}^{\Lambda}
+\Sigma_{12\downarrow\downarrow}^{\Lambda}) \label{selfrg} \\
&&\frac{d(\Sigma_{12\uparrow\uparrow}^{\Lambda}
-\Sigma_{12\downarrow\downarrow}^{\Lambda})}{d l}
=2b(g_{1s}+g_{1t})
(\Sigma_{12\uparrow\uparrow}^{\Lambda}
-\Sigma_{12\downarrow\downarrow}^{\Lambda}) \\
&&\frac{d\Sigma_{12\uparrow\downarrow}^{\Lambda}}{d l}
=2b(g_{1s}+g_{1t})\Sigma_{12\uparrow\downarrow}^{\Lambda}.
\end{eqnarray}
Because the strongest divergence of the 4-point vertex appears 
in $3g_{1t}-g_{1s}$ (see FIG.3), the off-diagonal
self-energy $\sum_{\sigma}\Sigma_{12\sigma\sigma}$
becomes non-zero at some critical $\Lambda=\Lambda_c$.
This is easily seen by solving (\ref{selfrg}), which gives
\begin{eqnarray}
\sum_{\sigma}\Sigma^{\Lambda}_{12\sigma\sigma}=
\sum_{\sigma}\Sigma^{\Lambda_0}_{12\sigma\sigma}
{\rm exp}[2b\int^l_0 dl'(3g_{1t}-g_{1s})].
\end{eqnarray}
Although $\Sigma_{12\sigma\sigma}^{\Lambda_0}$ is vanishing
in the vicinity of the $\Gamma$ point, because of the momentum dependence
of $s_{\mu\nu}(k)$, 
$\sum_{\sigma}\Sigma_{12\sigma\sigma}^{\Lambda_c}$ becomes non-zero
for $\Lambda=\Lambda_c$ at which value
$3g_{1t}-g_{1s}$ is divergent.
The non-zero off-diagonal self-energy hybridizes the band 1 and the band 2
at the $\Gamma$ point, and drives the system into 
{\it the insulating state with both spin and charge gaps.}
Thus the singularity of the RG flow signifies the (semi-)metal-to-insulator
transition.

\subsection{Mean Field Analysis}

The above RG analysis implies the existence of a mean field solution
for which the order parameter is given by,
\begin{eqnarray}
\Delta_k\equiv
\sum_{\sigma}\Sigma_{12\sigma\sigma}(k)
\sim \sum_{\sigma={\uparrow\downarrow}}
\langle a_{k 1\sigma}^{\dagger}a_{k 2\sigma}\rangle.
\end{eqnarray}
This state is characterized by electron-hole pairing with parallel spins, 
which leads to the formation of both spin and charge gaps
preserving the spin rotational symmetry. 
$\Delta_k$ is determined by the self-consistent mean field equation,
which is obtained as follows.
According to the numerical analysis of the RG equations
(\ref{rge1})-(\ref{rge4}),
$g_{4s}$ is mainly renormalized by the first term
of the right-hand side of (\ref{rge3}).
Then, the renormalized coupling $g_{4s}$ is 
approximately given by RPA-like expressions.
As a result,  
the self-consistent gap equation for $\Delta_k$ is 
expressed diagrammatically as shown in FIG.4.
The fist term of the right-hand side of the gap equation
gives less singular contributions than the second term. 
Thus, the linearized gap equation, which determines 
the transition temperature, is, 
\begin{eqnarray}
\Delta_k&=&\sum_{q,k'}\Pi(k,q-k)
G_{11}(q-k)G_{22}(q-k) \nonumber \\
&\times&\Pi(k',q-k)G_{11}(k')G_{22}(k')\Delta_{k'}, \label{gap}
\end{eqnarray}
where $G_{\mu\mu}(k)=1/(\varepsilon_n-E_{k\mu})$,  
$k=({\rm i}\varepsilon,\mbox{\boldmath $k$})$, and,
\begin{eqnarray}
\Pi(k,k')&=&\sum_{\nu=\pm}\frac{\nu
Ut^{\nu}(\mbox{\boldmath $k$},\mbox{\boldmath $k'$)}}
{1-c_{\nu}D(k+k')}, \\  
t^{\pm}(\mbox{\boldmath $k$},\mbox{\boldmath $k'$})
&=&(s_{11}(\mbox{\boldmath $k$})s_{12}(\mbox{\boldmath $k'$})
\pm s_{12}(\mbox{\boldmath $k$})s_{11}(\mbox{\boldmath $k'$}))^2/2, \\
D(q)&=&-TU\sum_{n,k} G_{11}(k)G_{11}(q-k), 
\end{eqnarray}
with $c_{+}=2a$, and $c_{-}=\sum_k s_{11}^2$.
Here we have ignored the diagonal self-energy. 
Equation (\ref{gap}) implies that the gap function can be written
\begin{eqnarray}
\Delta_k=s_{11}(k)s_{12}(k)\Delta_0
=\frac{\Delta_0(\cos k_x-\cos k_y)}{2(1-\cos k_x\cos k_y)}, \label{delta}
\end{eqnarray} 
where $\Delta_0$ is a constant. From (\ref{gap}), we have
the equation that determines the transition temperature,
\begin{eqnarray}
1=\frac{U^2}{16t}
[\ln\left(\frac{8t}{U}\right)-b_0]\ln\left(\frac{8t}{T_c}\right),
\label{tran}
\end{eqnarray}
where $b_0=0.322$. (\ref{tran}) implies that 
for $U<U_c\sim 0.725\times 8t$,
a state with non-zero $\Delta_0$ is realized.
Note that the gap function (\ref{delta}) has a line node structure similar to
$d_{x^2-y^2}$ symmetry.
However, this line node vanishes, when we take into account the coupling 
with lattice, as will be discussed in Sec. IV.D.

Generally, in 2D systems, thermal fluctuations may
suppress the transition temperature down to zero. 
The above mean field solution is also affected seriously by
thermal fluctuations, because U(1) Goldstone mode related with the
phase of $\langle a_{k 1\sigma}^{\dagger}a_{k 2\sigma}\rangle$
does not survive at finite temperatures. 
To see this, we have applied the Ginzburg-Landau analysis to 
this mean field solution and found that, in the 2D case, the transition
temperature vanishes in accordance with the Mermin-Wagner-Coleman theorem.
Nevertheless, the above analysis demonstrates that in the ground state 
at zero temperature, the gap $\Delta_k$ is non-zero, 
and the system is in an insulating state.

\begin{figure}[h]
\centerline{\epsfxsize=5.5cm \epsfbox{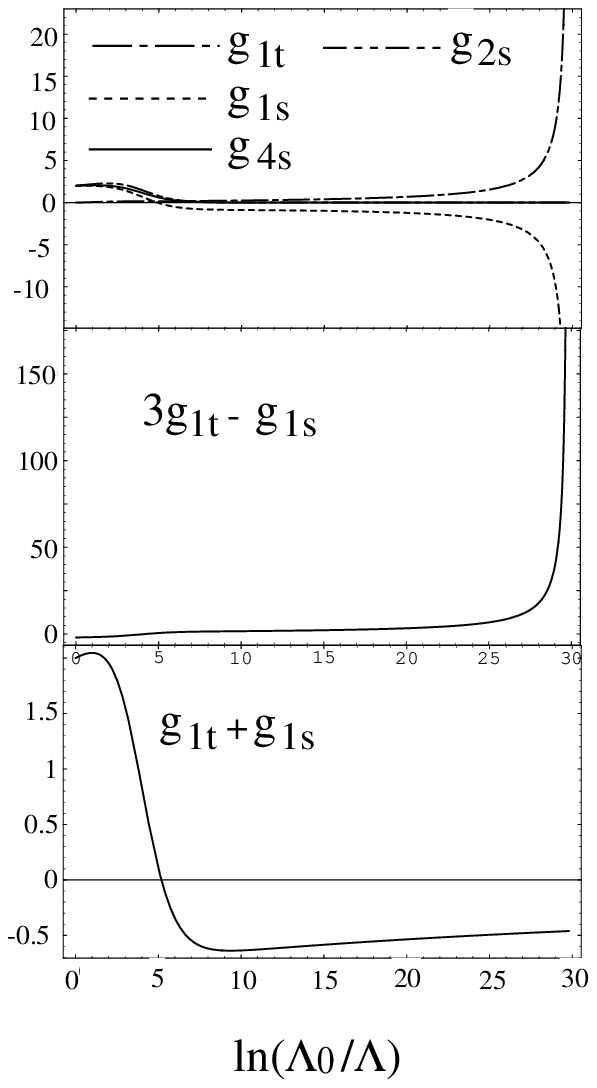}}
{FIG.3 The RG flow of the running couplings in the 2D case with 
$U/8t=0.25$.}
\end{figure}

\subsection{Properties of the Insulating state}

\subsubsection{Spin Gap State with Spin Rotational Symmetry}

We now further investigate the properties of the insulating phase
using the mean field solution.
The single-particle
Green's functions in this state are given by,
\begin{eqnarray}
&&G^{\rm MF}_{11}(k,\varepsilon_n)
=\frac{a^{(+)}}{{\rm i}\varepsilon+\mu-E_k^{(+)}}
+\frac{a^{(-)}}{{\rm i}\varepsilon+\mu-E_k^{(-)}}, \\
&&G^{\rm MF}_{22}(k,\varepsilon_n)=\frac{a^{(-)}}
{{\rm i}\varepsilon+\mu-E_k^{(+)}}
+\frac{a^{(+)}}{{\rm i}\varepsilon+\mu-E_k^{(-)}},
\end{eqnarray}
where 
\begin{eqnarray}
E_{k}^{(\pm)}&=&(E_{k1}+E_{k2}\pm
\sqrt{(E_{k1}-E_{k2})^2+4\Delta_k^2})/2, \\ 
a^{(\pm)}&=&\pm \frac{E^{(\pm)}-E_{k1}}{E_k^{(+)}-E_k^{(-)}}.
\end{eqnarray}
In the insulating phase, because the order parameter 
does not break the spin rotational symmetry,
there is no long-range magnetic order. 
However, a spin gap exists.
The spin gap behavior is observed in the
temperature dependence of the spin-lattice relaxation rate $1/T_1$ 
probed by NMR measurements.
It is obtained 
from the above mean field solution,
\begin{eqnarray} 
\frac{1}{T_1T}\sim \int dE \frac{[N_i(E)]^2}{2T\cosh^2 \frac{E}{2T}}, \\
N_i(E)=\Bigl\langle 
\frac{\sqrt{E^2-\Delta_k^2}}{E^2+\Delta_k^2}\Bigr\rangle_{k\sim 0} 
E^{\frac{3}{2}}. 
\label{spinco}
\end{eqnarray}
Here, $\langle\cdot\cdot\cdot\rangle_{k\sim 0}$ is
the angular average near $k=0$.
Because of the nodes of $\Delta_k$, we have, $1/(T_1T)\sim T^3$.
However, as will be discussed in the next section,
the coupling with lattice degrees of freedom changes this power law behavior
to an exponential decay, $1/(T_1T)\sim\exp(-\Delta/T)$.

\subsubsection{Charge Ordering}

Another important property of the insulating phase manifests in 
the charge degrees of freedom.
The formation of the gap $\Delta_k$ brings about a difference between
the charge densities at the sites 1 and 2 in a unit cell given by
\begin{eqnarray}
\rho_1-\rho_2&=&-4\sum_k \frac{s_{11}(k)s_{12}(k)\Delta_k}
{\sqrt{(E_{k2}-E_{k1})^2+4\Delta_k^2}}\Theta(\mu-E_k^{(-)}) \nonumber \\
&&\sim\Delta_0/t. \label{cha}
\end{eqnarray}
Thus charge ordering (CO) with a charge density displacement
proportional to the gap characterizes this insulating state. 
The CO pattern is shown in FIG.5.
This noteworthy result can be understood as follows.
In our system, three electrons occupying nearest neighbor sites
cost energy loss caused by magnetic frustration.
Conversely, magnetic frustration induces 
an effective finite-range repulsion 
between electrons at nearest neighbor sites.
If this finite-range repulsion is sufficiently strong to
overcomes the on-site Coulomb interaction $U$,
the CO state will be stabilized. 
This is possible if $U$ is not so large. 
As $U$ increases, a transition to a conventional Mott insulator 
with no charge ordering should occur. 
This transition cannot be described within our weak-coupling analysis.  

The CO pattern shown in FIG.5 is regarded as
an assembly of one-dimensional chains in the $[1,1]$ and $[1,-1]$ directions.
This observation implies that CO reduces the spatial dimension effectively
to relax geometrical frustration.

\begin{figure}[h]
\centerline{\epsfxsize=7.8cm \epsfbox{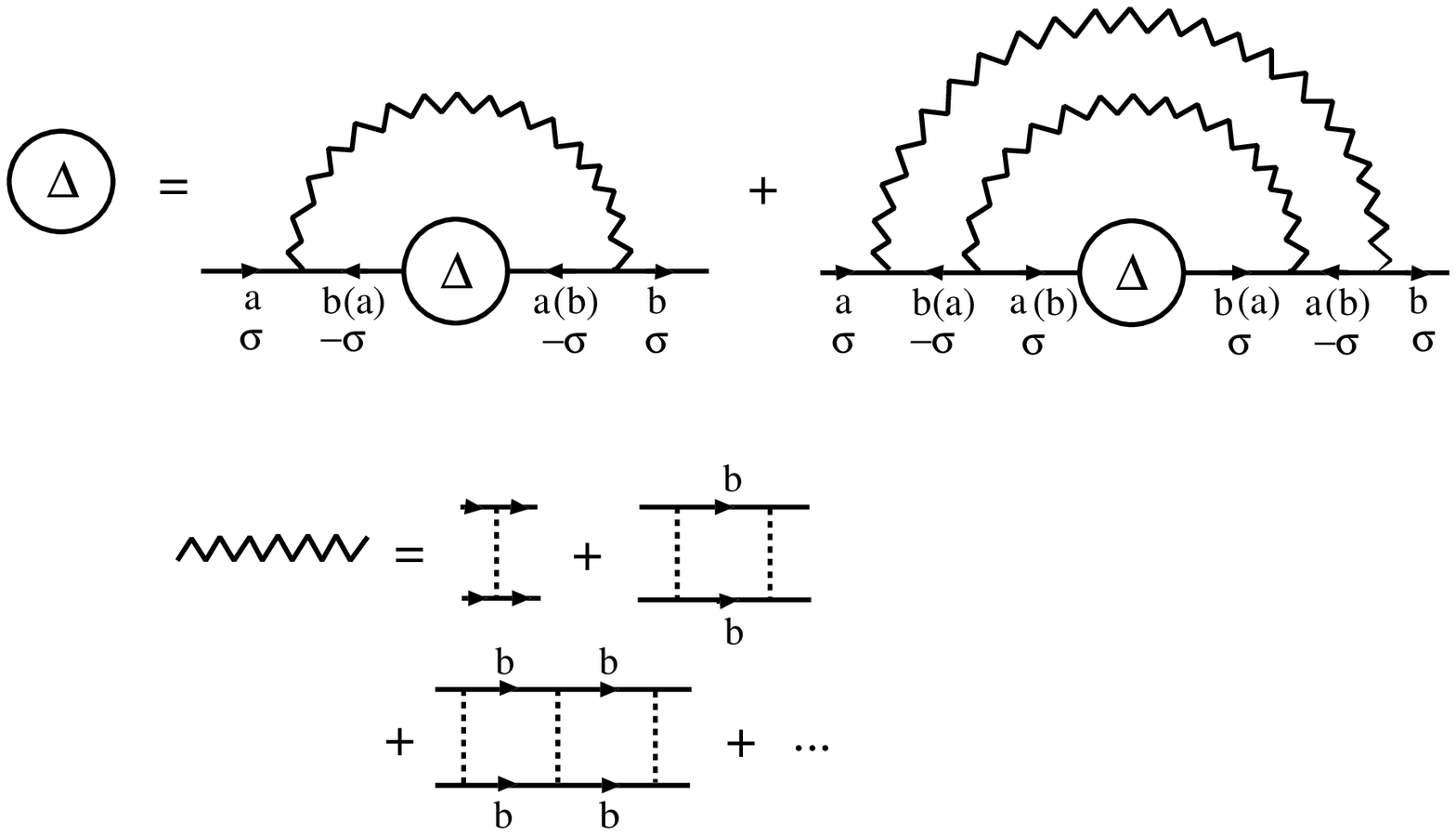}}
{FIG.4 Diagrams for the linearized gap equations.}
\end{figure}

\begin{figure}[h]
\centerline{\epsfxsize=2.4cm \epsfbox{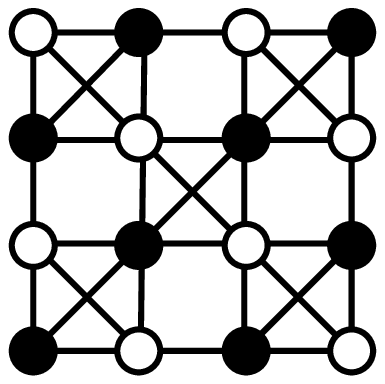}}
{FIG.5 The CO pattern in the 2D case.}
\end{figure}

\subsection{Coupling with Lattice Distortion}

The presence of CO found in the
previous section implies that the coupling between the charge fluctuation and
lattice degrees of freedom may give rise to the distortion of the lattice
structure.
Since the band structure possesses the double degeneracy 
at the $\Gamma$ point even for $U=0$ in the half-filling case,
the Jahn-Tellar distortion may occur to lift the degeneracy.
However, in our system the lift of the double degeneracy
corresponds to the situation that the charge density on the site 1 is
different from that on the site 2.  
It is highly non-trivial whether the inhomogeneous charge
distribution is realized even in the presence of the on-site repulsion $U$,
which, in general, should suppress the charge fluctuation.
The results obtained in the previous sections remarkably show 
that the interplay between the one-site repulsion and geometrical frustration  
stabilizes the charge ordering in the ground state, to which 
the Jahn-Tellar lattice distortion may adjust.
In this section, we discuss the distortion of the lattice structure
compatible with the CO state.

Since the band degeneracy appears only at the $\Gamma$ point, it is sufficient
to consider the point group of the tetragonal crystal system ${\rm D_{4h}}$.
The doubly degenerated levels at the $\Gamma$ point belong to 
${\rm E_u}$ representation. 
There are three normal modes which are relevant to the Jahn-Tellar
distortion: $[{\rm E_u^2}]={\rm A_1}+{\rm B_{1g}}+{\rm B_{2g}}$.
The normal coordinates of these modes are, respectively, given by,
\begin{eqnarray}
Q({\rm A_{1g}})&=&-u_{1x}-u_{1y}+u_{2x}-u_{2y} \nonumber \\
&&+u_{3x}+u_{3y}-u_{4x}+u_{4y}, \\
Q({\rm B_{1g}})&=&u_{1x}+u_{1y}+u_{2x}-u_{2y} \nonumber \\
&&-u_{3x}-u_{3y}-u_{4x}+u_{4y}, \\
Q({\rm B_{2g}})&=&-u_{1x}+u_{1y}+u_{2x}+u_{2y} \nonumber \\
&&+u_{3x}-u_{3y}-u_{4x}-u_{4y}, 
\end{eqnarray}  
where $u_{i\alpha}$ ($i=1,2,3,4$ and $\alpha=x,y,z$) is 
the displacement of the $i$-site in the $\alpha$-direction.
The lattice distortions corresponding to these modes are schematically shown
in FIG.6. 
The lattice distortions change the kinetic term of the Hamiltonian
$H_{\rm kin}\rightarrow H_{\rm kin}+\Delta H_{\rm kin}$ with,
\begin{eqnarray}
\Delta H_{\rm kin}&=&-a_{\rm A_{1g}}\sum_{i,j}(c^{\dagger}_{1i}c_{1j}
+c^{\dagger}_{2i}c_{2j}  \nonumber \\
&&+c^{\dagger}_{1i}c_{2j}+c^{\dagger}_{2i}c_{1j})Q({\rm A_{1g}}) \nonumber \\
&&-a_{\rm B_{1g}}\sum_{i,j}(c^{\dagger}_{1i}c_{1j}-c^{\dagger}_{2i}c_{j})
Q({\rm B_{1g}}) \nonumber \\
&&-a_{\rm B_{2g}}
\sum_i(c^{\dagger}_{1i}c_{2i+a_x}-c^{\dagger}_{1i}c_{2i+a_y} \nonumber \\
&&+c^{\dagger}_{2i}c_{1i+a_x}-c^{\dagger}_{2i}c_{1i+a_y}+h.c.)Q({\rm B_{2g}}).
\end{eqnarray}
Here we omit the spin index.
The ${\rm A_{1g}}$ and ${\rm B_{2g}}$ modes do not lift the degeneracy
at the $\Gamma$ point.
Obviously, the ${\rm B_{1g}}$ mode is the only lattice distortion that
is consistent with the CO pattern shown in FIG.5.
After this lattice distortion occurs, the single-particle excitation gap
completely opens at the $\Gamma$ point without line nodes, 
leading the exponential decay of
the spin lattice relaxation rate, as announced in the previous section.

\begin{figure}[h]
\centerline{\epsfxsize=7cm \epsfbox{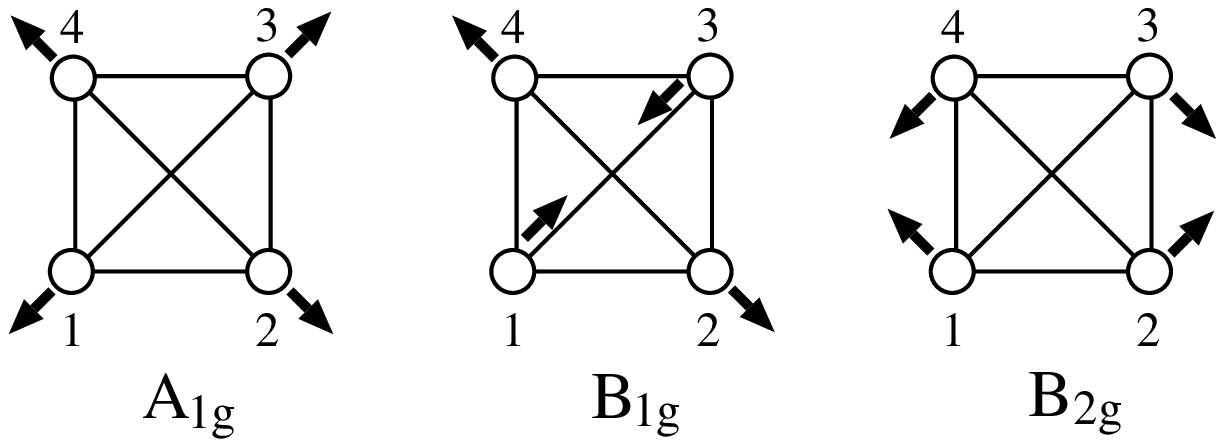}}
{FIG.6. Three normal modes of the lattice distortion.}
\end{figure}

\section{3D Pyrochlore Hubbard Model}

\subsection{RG Analysis}

The above analysis can be straightforwardly 
generalized to the case of a 3D pyrochlore lattice.
We obtain the RG flow numerically from
(\ref{rge1})-(\ref{rge10}) for $d=3$.
Here, in contrast to the 2D case, for sufficiently small $U$
all couplings are irrelevant, and thus the semi-metal state is stable.
However, for values of $U$ larger than a certain critical value 
but still smaller than the band width,
RG flow similar to that in the 2D case is obtained, as shown in FIG.7.
The coupling $3g_{1t}-g_{1s}$, which is related to
the charge degrees of freedom, scales into the strong-coupling regime.
This RG flow implies that, as in the 2D case, 
a particle-hole pairing between the flat bands and 
the dispersive band is realized below a critical temperature,
leading spontaneous gap formation both in the charge and spin degrees of
freedom.
Note that the spin gap formation does not signify the breaking of 
the spin rotational symmetry,
since the vertex in the spin degrees of freedom $g_{1s}+g_{1t}$ is
still finite at the critical value of $\ln(\Lambda_0/\Lambda)$.

\subsection{Mean Field Analysis}

Although the value of $U$ used above is relatively large, 
we expect that the one-loop RG calculation still gives qualitatively
correct results, as long as $U$ is smaller than the band width.
To examine the validity of the one-loop calculation, 
we explore the self-consistent mean field solution.  
The order parameters for the particle-hole pairing state 
suggested from the above RG flows is, 
\begin{eqnarray}
\Delta^{(13)}_k&=&\sum_{\sigma}
\langle a^{\dagger}_{k1\sigma}a_{k3\sigma}\rangle, \label{op1} \\
\Delta^{(23)}_k&=&\sum_{\sigma}
\langle a^{\dagger}_{k2\sigma}a_{k3\sigma}\rangle.  \label{op2}
\end{eqnarray}
The self-consistent gap equations for the 3D case are
also given by the diagram shown in FIG.4, from which we find that
the gap functions are given by,
\begin{eqnarray}  
\Delta^{(13)}_k&=&\sum_{\nu=1}^4 s_{\nu 1}(k)s_{\nu 3}(k)
\Delta_{\nu}^{(13)} \nonumber \\
&=&\sum_{\nu=1}^2s_{\nu 1}(k)s_{\nu 3}(k)
(\Delta_{\nu}^{(13)}-\Delta_3^{(13)}), \\
\Delta^{(23)}_k&=&\sum_{\nu=1}^4 s_{\nu 2}(k)s_{\nu 3}(k)
\Delta_{\nu}^{(23)} \nonumber \\
&=&\sum_{\nu=1}^3s_{\nu 2}(k)s_{\nu 3}(k)
(\Delta_{\nu}^{(23)}-\Delta_4^{(23)}).
\end{eqnarray}
Here we used the orthogonal relations, $\sum_{\nu=1}^4s_{\nu 1}s_{\nu 3}=0$,
$\sum_{\nu=1}^4s_{\nu 2}s_{\nu 3}=0$, and $s_{41}=0$.
Using the symmetry properties of $s_{\mu\nu}(k)$ in momentum space,
we can impose some restrictions on the structure of the gap functions
without solving the gap equations.
Under the transformation, $x\rightarrow y$, 
$y\rightarrow x$, $z\rightarrow -z$, the coefficients of $\Delta^{(13)}_k$
are transformed as,
\begin{eqnarray} 
&&s_{11}s_{13}\rightarrow -s_{21}s_{23},\\
&&s_{21}s_{23}\rightarrow -s_{11}s_{13}, \\
&&s_{31}s_{33}\rightarrow -s_{31}s_{33}. 
\end{eqnarray}
Because of the symmetry of a tetrahedron ${\rm T_d}$, 
the gap function should be 
unchanged up to the sign by this transformation. 
Then, we have, 
\begin{eqnarray}
&&\Delta^{(13)}_1=\Delta^{(13)}_2,  \quad \mbox{or} \\
&&\Delta^{(13)}_1+\Delta^{(13)}_2=2\Delta^{(13)}_3. \label{rel1}
\end{eqnarray}
In a similar manner, using the transformation,
$x\rightarrow x$, $y\rightarrow z$, $z\rightarrow y$,  we obtain,
\begin{eqnarray}
&\Delta^{(13)}_1=\Delta^{(13)}_3, \quad \mbox{or} \\
&\Delta^{(13)}_1+\Delta^{(13)}_3=2\Delta^{(13)}_2. \label{rel2}
\end{eqnarray}
Combining (\ref{rel1}) and (\ref{rel2}), we end up with 
$\Delta^{(13)}_1=\Delta^{(13)}_2=\Delta^{(13)}_3$, and thus,  
\begin{equation}
\Delta^{(13)}_k=0.
\end{equation}
Applying a similar argument to $\Delta^{(23)}_k$, we find
$\Delta^{(23)}_1=\Delta^{(23)}_2=\Delta^{(23)}_3$, and,
\begin{equation}
\Delta^{(23)}_k=s_{42}(k)s_{43}(k)(\Delta_4^{(23)}-\Delta_1^{(23)}).
\end{equation} 
The quantity $\Delta_4^{(23)}-\Delta_1^{(23)}$ 
is determined from the gap equation.
According to the RG analysis, the transition occurs only for sufficiently
large $U$. Therefore, to determine the transition temperature and 
the gap function correctly,
we need to take into account 
the self-energy corrections i.e., pair breaking effect.
This calculation is rather involved, and we have not yet carried it out.
However, we see from the RG flow that at the critical temperature 
$T_c\sim \Lambda_0 e^{-l_c}=8t\times 0.0042$, 
a transition from a semi-metal to an insulator occurs.
In the resulting insulating state, the three-fold degeneracy at
the $\Gamma$ point in the semi-metal state is lifted completely, and
a spin gap as well as a charge gap exists.
The gap function $\Delta^{(23)}_k$ has both line and point nodes determined by
$s_{42}(k)s_{43}(k)=0$.
However, these nodes are eliminated by the coupling with lattice degrees of
freedom,  as will be shown in Sec.V.D.

The particle-hole pairing state characterized by the order parameters
(\ref{op1}) and (\ref{op2}) is analogous to 
the excitonic insulator.\cite{ex}  
However, this analogy is not complete.
In contrast to the excitonic insulator which is realized in particle-hole
symmetric bands, the pairing state found above is stabilized by
strong interaction between flat bands and dispersive bands
in the absence of particle-hole symmetry.
Thus, we believe that this is a novel mechanism for the particle-hole pairing 
specific to pyrochlore systems.

\begin{figure}[h]
\centerline{\epsfxsize=5.5cm \epsfbox{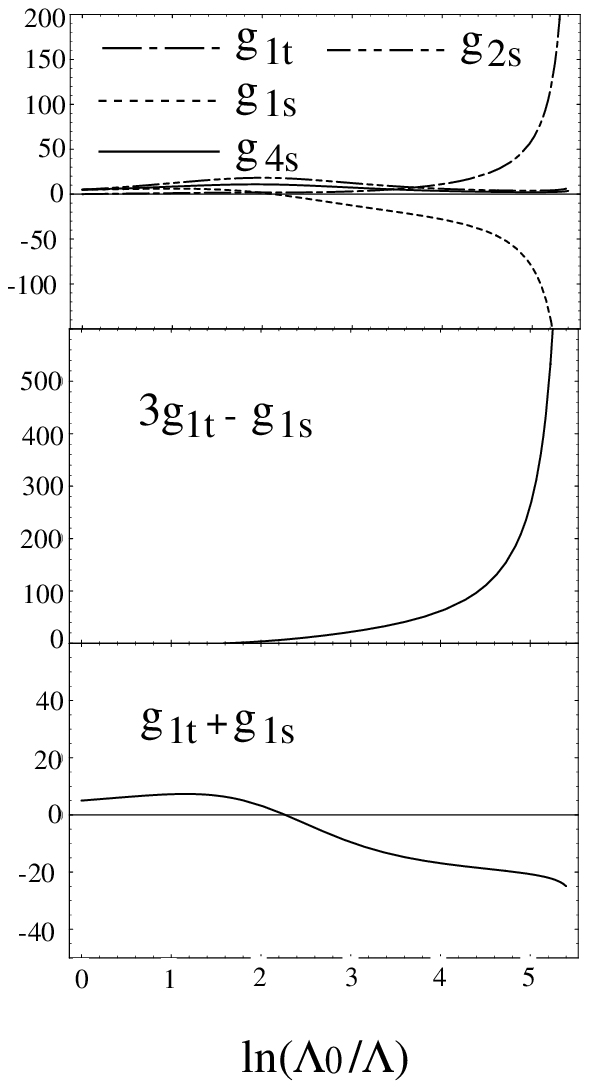}}
{FIG.7 The RG flow of the running couplings in the 3D case with 
$U/8t=0.625$.}
\end{figure}

\subsection{Properties of the Insulating State}

\subsubsection{Spin Gap State with Spin Rotational Symmetry}

It is straightforward to compute magnetic properties of the gapped state
using the mean field solution obtained in the previous section.
In the insulating phase with the single-particle excitation gap 
$\Delta^{(23)}_k$, the spin-lattice relaxation rate $1/T_1$
is also given by
(\ref{spinco}).
Since the gap function has nodes, $1/(T_1T)$ decreases in a power law
as temperature is lowered: $1/(T_1T)\sim T^3$.
However, as will be discussed in Sec. V.D,
in the presence of the coupling with lattice degrees of freedom,
the nodes of the gap function disappear. 
In this case, the spin-lattice relaxation rate shows a exponential 
behavior, $1/(T_1T)\sim \exp(-\Delta/T)$.

The spin gap behavior appears in the spin-spin correlation function 
${\rm Im}\chi_s(q,\omega)$ for all $q$.
The momentum dependence of ${\rm Im}\chi_s(q,\omega)$ is quite
small, indicating that geometrical frustration suppresses any tendency
toward a conventional magnetic order. 
Such small $q$-dependence of ${\rm Im}\chi_s(q,\omega)$
was also found in the semi-metal state
above $T_c$.\cite{fuji}

\subsubsection{Charge Ordering}

Here, we examine the possibility of a CO state
in the 3D system. There are four sites in a unit cell.
The appearance of a gap causes a charge density displacement
on each site given by,
\begin{eqnarray}
\delta\rho_{\nu}=2\sum_ks_{\nu 2}(k)s_{\nu 3}(k)\Delta^{(23)}_k,
\end{eqnarray}
for $\nu=1,2,3$, and,
\begin{eqnarray}
\delta\rho_4=-\sum_{\nu=1}^3\delta\rho_{\nu}.
\end{eqnarray}
Using the symmetry properties of $s_{\mu\nu}(k)$, we obtain,
\begin{eqnarray}
\delta\rho_1&=&\delta\rho_2=\delta\rho_3\neq 0, \\ 
\delta\rho_4&=&-3\delta\rho_1.
\end{eqnarray} 
It is thus found that CO with the pattern displayed in FIG.8 
occurs in the insulating phase.
Interestingly, a similar CO pattern is observed in
the spinel system ${\rm AlV_2O_4}$ which possesses a V-site 
corner-sharing tetrahedron network.\cite{al}
Within the above analysis, the sign of $\delta\rho_4$, which depends on
the phase of the order parameter $\Delta^{(23)}_4-\Delta^{(23)}_1$,
is not determined.
Thus, the states of $\delta\rho_4>0$ and $\delta\rho_4<0$ are degenerated.
As will be discussed in the next section, 
this degeneracy is lifted by the coupling with lattice degrees of freedom.

Note that the CO pattern shown in FIG.8 
is regarded as an alternate stack of 2D kagome
lattices and triangular lattices in the $[-1,1,-1]$ direction.
As in the 2D case, CO brings about
the effective reduction of spatial dimension 
to relax geometrical frustration.   

\begin{figure}[h]
\centerline{\epsfxsize=2.6cm \epsfbox{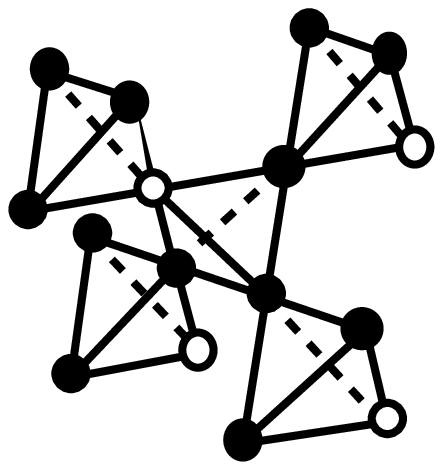}}
{FIG.8. The CO pattern in
the 3D case.}
\end{figure}

\subsection{Coupling with Lattice Distortion}

As in the 2D case, CO found in the previous section may bring about
lattice distortion.
Here, we discuss this possibility. 
Since, at the half-filling, all important processes occur in the vicinity of
the $\Gamma$ point, it is sufficient to consider the point group 
of the tetrahedron ${\rm T_d}$. The representation of ${\rm T_d}$ is reduced
to ${\rm A_1}+{\rm T_2}$. The triply degenerated levels at the $\Gamma$
point $E_{k1}=E_{k2}=E_{k3}$ ($k=0$) belong to ${\rm T_2}$.
Thus, the normal modes that may lift the degeneracy are obtained
from the symmetric product representation, 
$[{\rm T_2}^2]={\rm A_1}+{\rm E}+{\rm T_2}$.
The normal coordinates corresponding to these three modes are,
\begin{eqnarray}
Q({\rm A_1})&=&(u_{1x}-u_{2x}+u_{3x}-u_{4x} \nonumber \\
&&+u_{1y}-u_{2y}-u_{3y}+u_{4y} \nonumber \\
&&+u_{1z}+u_{2z}-u_{3z}-u_{4z})/\sqrt{3}, \\
Q^{(1)}({\rm E})&=&(u_{1x}-u_{2x}+u_{3x}-u_{4x}\nonumber \\
&&+u_{1y}-u_{2y}-u_{3y}+u_{4y}\nonumber \\
&&-2u_{1z}-2u_{2z}+2u_{3z}+2u_{4z})/\sqrt{6}, \\
Q^{(2)}({\rm E})&=&(u_{1x}-u_{2x}+u_{3x}-u_{4x} \nonumber \\
&&-u_{1y}+u_{2x}+u_{3x}-u_{4x})/\sqrt{2},  \\
Q^{(1)}({\rm T_2})&=&u_{1x}+u_{2x}-u_{3x}-u_{4x}, \\
Q^{(2)}({\rm T_2})&=&u_{1y}-u_{2y}+u_{3y}-u_{4y}, \\
Q^{(3)}({\rm T_2})&=&u_{1z}-u_{2z}-u_{3z}+u_{4z}.
\end{eqnarray}
They are schematically shown in FIG.9(a), (b), and (c).
The breezing mode ${\rm A_1}$ does not lift the degeneracy 
at the $\Gamma$ point.
The ${\rm E}$ and ${\rm T_2}$ modes split the triply degenerated levels
into three different levels or one separated level and doubly degenerated
levels.
The modification of the kinetic energy due to these modes are,
\begin{eqnarray}
\Delta H_{\rm kin}&=&b_{\rm E}\sum_{i,j}\Bigl[
\frac{Q^{(1)}({\rm E})}{2\sqrt{3}}
(2c^{\dagger}_{i1}
c_{j2}+2c^{\dagger}_{i3}c_{j4}-c^{\dagger}_{i1}c_{j3}  \nonumber \\
&-&c^{\dagger}_{i2}c_{j3}-c^{\dagger}_{i2}c_{j4}
-c^{\dagger}_{i1}c_{j4}+h.c.) \nonumber \\
&+&\frac{Q^{(2)}({\rm E})}{2}(c^{\dagger}_{i1}c_{j4}+c^{\dagger}_{i2}c_{j3}
-c^{\dagger}_{i1}c_{j3}-c^{\dagger}_{i2}c_{j4}+h.c.)\Bigr] \nonumber \\
&+&b_{\rm T_2}\sum_{i,j}\Bigl[\frac{Q^{(1)}({\rm T_2})}{\sqrt{2}}
(c^{\dagger}_{i1}c_{j4}-c^{\dagger}_{i2}c_{j3}+h.c.) \nonumber \\
&+&\frac{Q^{(2)}({\rm T_2})}{\sqrt{2}}
(c^{\dagger}_{i1}c_{j2}-c^{\dagger}_{i3}c_{j4}+h.c.) \nonumber \\
&+&\frac{Q^{(3)}({\rm T_2})}{\sqrt{2}}
(c^{\dagger}_{i1}c_{j3}-c^{\dagger}_{i2}c_{j4}+h.c.)\Bigr].
\end{eqnarray}
Since the stretch of bonds may reduce the hopping integral,
it is plausible to assume $b_{\rm E}>0$ and $b_{\rm T_2}>0$. 
The lattice distortion that is consistent with 
the CO pattern shown in FIG.8 is obtained by putting 
$Q^{(1)}({\rm T_2})=-Q^{(2)}({\rm T_2})=-Q^{(3)}({\rm T_2})\neq 0$, and
$Q^{(1)}({\rm E})=Q^{(2)}({\rm E})=0$.
This mode is shown in FIG.10. Here we have dropped
the rotational degrees of freedom around $(-1,1,-1)$-axis.
For $Q^{(1)}({\rm T_2})>0$, the triply degenerated levels at the $\Gamma$
point split into one lower level and upper doubly degenerated levels.
Thus, at the half-filling, 
this lattice distortion generates a full gap without a node
in the single-particle excitation.
Note that the full gap opens not only at the $\Gamma$ point but also
over the entire Brillouin zone, as easily verified by
diagonalizing the kinetic term with this lattice distortion.
On the other hand, for $Q^{(1)}({\rm T_2})<0$, the lower levels
are doubly degenerated, and the system is still in the semi-metallic state
unless we do not take into account the correlation-driven 
CO transition obtained in the previous sections.
Thus it is expected that the case of $Q^{(1)}({\rm T_2})>0$, in which
both the lattice distortion and electron correlation stabilize 
the gapped state, is energetically favorable.
Under the trigonal distortion with $Q^{(1)}({\rm T_2})>0$ shown in FIG.10, 
the charge density on the site 4 in the tetrahedron is larger 
than those on the other three sites.  
This is easily seen by calculating straightforwardly
the charge density on each site 
in a single tetrahedron with the trigonal distortion. 
Then, the sigh of $\delta\rho_4$, which is not determined in the absence of
the lattice-coupling as mentioned in the previous sections,
is chosen as $\delta\rho_4>0$ by the lattice distortion.

The lattice distortion discussed here and the CO state shown in FIG.8
are similar to the experimental
observation for ${\rm AlV_2O_4}$, apart from the doubling of the unit cell
in the $[1,1,1]$ direction found in ${\rm AlV_2O_4}$.\cite{al}
Since, in ${\rm AlV_2O_4}$, the $t_{2g}$ orbital of V sites
plays a central role, our simple model is not directly applicable to
this system.
However, according to the recent LDA calculation, the band structure near
the Fermi surface possesses $s$-like character 
because the $s$ orbital of Al site
partially hybridizes with the $a_{1g}$ orbital split from the $t_{2g}$
orbital by a trigonal crystal field.\cite{harima}
Thus, CO observed in ${\rm AlV_2O_4}$ may be
microscopically explained by the mechanism described here.
It should be emphasized that in our scenario, 
the interplay between electron correlation and
geometrical frustration is the most important ingredient for the realization
of the CO state and the lattice distortion is merely a secondary effect. 

\begin{figure}[h]
\centerline{\epsfxsize=7.5cm \epsfbox{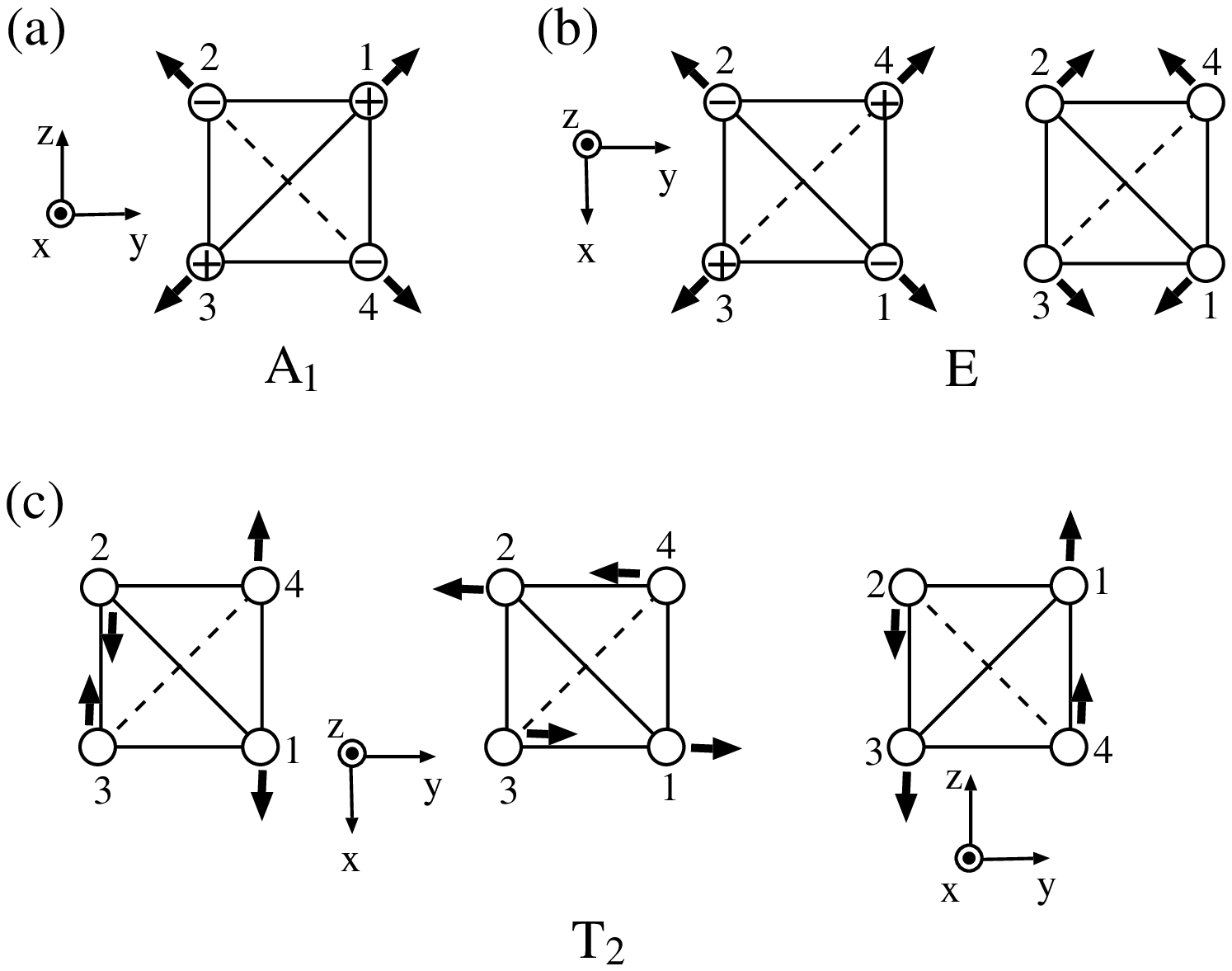}}
{FIG.9. The normal modes of ${\rm T_d}$. (a) ${\rm A_1}$ (b) ${\rm E}$ 
(c) ${\rm T_2}$.}
\end{figure}

\begin{figure}[h]
\centerline{\epsfxsize=3cm \epsfbox{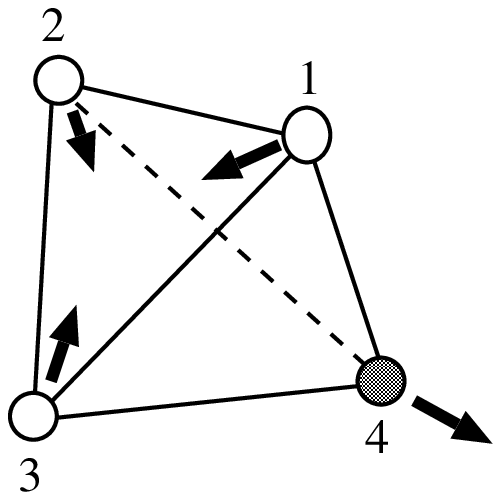}}
{FIG.10. The lattice distortion consistent with the CO pattern shown in FIG.8.}
\end{figure}

\subsection{Application to MIT in ${\rm Tl_2Ru_2O_7}$ }

We now apply the results obtained above 
to the description of the MIT observed in ${\rm Tl_2Ru_2O_7}$.
As mentioned in the introduction, according to Ishii and Oguchi,
the electronic structure of this system consists of 
the $s$-orbital of Tl sites,  as well as
the $t_{2g}$-orbitals of Ru sites.\cite{ogu} 
The band structure formed by the former is well approximated by
our model.
The band calculation gives the band width of this system 
$8t \sim 2$eV.\cite{ogu}
Experimental data on the size of $U$ do not exists.
However, typically, the value of $U$ for transition metal oxides
is $\sim 2$eV. 
This gives us reason to believe that the analysis presented in this paper,
which suggests that the MIT occurs for
large $U\sim 8t$, can be applied to the description of the 
${\rm Tl_2Ru_2O_7}$ system.
The transition temperature estimated from the RG analysis is
$T_c\sim 98$K, which is almost comparable with 
the experimental values $100\sim 120$K.\cite{take}
A recent NMR measurement of Tl nuclei 
has revealed the presence of a spin gap
in the insulating state, which is consistent with our results.\cite{tl}
The possible existence of a CO state and 
large enhancement of charge fluctuations above $T_c$
predicted in our theory have not yet been investigated
experimentally. 
The experimental determination of whether a CO state exists
for ${\rm Tl_2Ru_2O_7}$ is a crucial test of this theory.
When there exists coupling to a lattice, 
CO should accompany lattice distortion.
As discussed in Sec.V.C, the CO pattern found in this study 
suggests that the cubic lattice symmetry is broken down 
to trigonal symmetry. 
Actually, it has been found experimentally that, in ${\rm Tl_2Ru_2O_7}$,
the lattice structure changes discontinuously at the MIT point.
This observation seems to suggest the presence of 
large charge fluctuations in this system.

\section{Summary}

We have studied the MIT caused by the interplay between geometrical
frustration and electron correlation. 
We have found, using the RG method and mean field analysis, that, 
the 2D and 3D pyrochlore Hubbard models at the half-filling
show the transition from semi-metal to spin-gapped insulator.
The transition occurs at a finite critical temperature $T_c$ in the 3D case,
and at $T=0$ in the 2D case.
In the insulating state, CO occurs concomitantly
so as to relax geometrical frustration.
The results obtained here are successfully applied 
to the description of the MIT
observed in the pyrochlore oxide ${\rm Tl_2Ru_2O_7}$, 
though it is a future issue to examine experimentally the presence of
the CO state in this system, as predicted from our theory.
The CO pattern found in this paper is also very similar to 
that observed in ${\rm AlV_2O_4}$.
The mechanism for CO in this system may be explained by
the scenario described in this paper, because the electronic structure
of ${\rm AlV_2O_4}$
possesses partially flat bands,\cite{harima} which is a crucial
ingredient for CO induced by geometrical frustration. 
It is a future issue to explore this possibility taking into account
the $t_{2g}$ orbitals of V sites.

The insulating state found in our systems
is characterized by particle-hole pairing.
In this sense, it is analogous to the excitonic insulator.
However, there are some important differences between them.
In contrast to the excitonic insulator, the spin gapped insulator
in the pyrochlore Hubbard models is stabilized by the presence of flat bands
originated from geometrical frustration.
Interestingly, Singh {\it et al.} pointed out a possibility of
the excitonic insulator realized in the pyrochlore 
oxides ${\rm Cd_2Os_2O_7}$.\cite{sin}
Since its electronic structure near the Fermi energy consists of 
the $t_{2g}$ manifolds, our model is not simply applicable to this system.
It is an intriguing issue to extend our analysis to
more realistic models with this electronic structure.

\acknowledgements{}
 
The author is grateful to K. Yamada and H. Harima 
for invalueable discussions.
This work was supported by a Grant-in-Aid from the Ministry
of Education, Science, and Culture, Japan.
   


\end{multicols}
                                                                    

\begin{references}
\bibitem{rami} A. P. Ramirez, Annu. Rev. Mater. Sci. {\bf 24}, 453 (1994).

\bibitem{pyroex}  M. Shiga, H. Wada, Y. Nakamura, J. Deportes, and
K. R. A. Ziebeck, J. Phys. Soc. Jpn. {\bf 57}, 3141 (1988);
M. Shiga, K. Fujisawa, and H. Wada,{\it ibid} {\bf 62}, 1329 (1993). 

\bibitem{gau} B. D. Gaulin, J. N. Reimers, T. E. Mason, J. E. Greedan,
and Z. Tun, Phys. Rev. Lett.{\bf 69}, 3244 (1992).

\bibitem{har} M. J. Harris, M. P. Zinkin, Z. Tun, B. M. Wanklyn,
and I. P. Swainson, Phys. Rev. Lett.{\bf 73}, 189 (1994).

\bibitem{maeno1} D. Yanagishima and Y. Maeno, J. Phys. Soc. Jpn.{\bf 70},
2880 (2001).

\bibitem{fuka} H. Fukazawa and Y. Maeno, Phys. Rev. B, to appear. 

\bibitem{ue} Y. J. Uemura, A. Keren, K. Kojima, L. P. Le, G. M. Luke, 
W. D. Wu, Y. Ajiro, T. Asano, Y. Kuriyama, M. Mekata, H. Kikuchi, 
and K. Kakurai, Phys. Rev. Lett. {\bf 73}, 3306 (1994).

\bibitem{col} R. Coldea, D. A. Tennant, A. M. Tsvelik, and Z. Tylczynski, 
Phys. Rev. Lett. {\bf 86}, 1335 (2001). 

\bibitem{py} R. Moessner and J. T. Chalker, Phys. Rev. Lett. {\bf 80},
2929 (1998).

\bibitem{re} J. N. Reimers, Phys. Rev. B{\bf 45}, 7287 (1992).

\bibitem{can} B. Canals and C. Lacroix, Phys. Rev. Lett. {\bf 80},
2933 (1998). 

\bibitem{ber} A. B. Harris, A. J. Berlinsky, and C. Bruder, 
J. Appl. Phys. {\bf 69}, 5200 (1991).

\bibitem{iso} M. Isoda and S. Mori, J. Phys. Soc. Jpn. {\bf 67}, 4022
(1998).  

\bibitem{ya} Y. Yamashita and K. Ueda, Phys. Rev. Lett. {\bf 85}, 4960
(2000).  

\bibitem{ko} A. Koga and N. Kawakami, Phys. Rev. B{\bf 63}, 144432 (2001).

\bibitem{ts} H. Tsunetsugu, J. Phys. Soc. Jpn, {\bf 70}, 640 (2001);
Phys. Rev. B{\bf 65}, 024415 (2001).

\bibitem{check} S. E. Palmer and J. T. Chalker, Phys. Rev. B{\bf 64}, 094412
(2001). 

\bibitem{lieb} E. H. Lieb and P. Schupp, Phys. Rev. Lett. {\bf 83}, 5362
(1999). 

\bibitem{sta} O. A. Starykh, and R. R. P. Singh, and G. C. Levine, 
Phys. Rev. Lett.{\bf 88}, 167203 (2002).

\bibitem{auer} E. Berg, E. Altman, and A. Auerbach, cond-mat/0206384.

\bibitem{fou} J.-B. Fouet, M. Mambrini, P. Sindzingre, and C. Lhuillier, 
cond-mat/0108070.

\bibitem{kag} J. T. Chalker,P. C. W. Holdsworth, and E. F. Shender, 
Phys. Rev. Lett.{\bf 68}, 855 (1992).
 
\bibitem{chan} I. Ritchey, P. Chandra, and P. Coleman, 
Phys. Rev. B{\bf 47}, 15342 (1993).

\bibitem{lech} P. Lecheminant, B. Bernu, C. Lhuillier, L. Pierre, 
and P. Sindzingre, Phys. Rev. B{\bf 56}, 2521 (1997).

\bibitem{tch} O. Tchernyshyov, R. Moessner, and S. L. Sondhi, 
Phys. Rev. Lett. {\bf 88}, 067203 (2002).

\bibitem{li} S. Kondo, D. C. Johnston, C. A. Swenson, F. Borsa, A. V. Mahajan, 
L. L. Miller, T. Gu, A. I. Goldman, M. B. Maple, D. A. Gajewski,
E. J. Freeman, N. R. Dilley, R. P. Dickey, J. Merrin, K. Kojima, G. M. Luke, 
Y. J. Uemura, O. Chmaissem, and J. D. Jorgensen, Phys. Rev. Lett.{\bf 78},
3729 (1997). 

\bibitem{fuji} S. Fujimoto, Phys. Rev. B{\bf 64}, 085102 (2001).

\bibitem{iso2} M. Isoda and S. Mori, J. Phys. Soc. Jpn. {\bf 69}, 
1509 (2000). 

\bibitem{fuji2} S. Fujimoto, Phys. Rev. B{\bf 65}, 155108 (2002).

\bibitem{ts2} H. Tsunetsugu, J. Phys. Soc. Jpn.{\bf 71}, 1844 (2002).

\bibitem{ya2} Y. Yamashita and K. Ueda, cond-mat/0212149.

\bibitem{cole} J. Hopkinson and P. Coleman, Phys. Rev. Lett. {\bf 89},
267201 (2002).

\bibitem{naga} R. Shindou and N. Nagaosa, Phys. Rev. Lett.{\bf 87}, 116801
(2001).

\bibitem{kawa} Y. Taguchi, Y. Oohara, H. Yoshizawa, N. Nagaosa, and
Y. Tokura, Science {\bf 291}, 2573 (2001).

\bibitem{take} T. Takeda, M. Nagata, H. Kobayashi, R. Kanno, Y. Kawamoto,
M. Takano, T. Kamiyama, F. Izumi, and A. W. Sleight,
J. Solid State Chem. {\bf 140}, 182 (1998).

\bibitem{mand} D. Mandrus, J. B. Thompson, R. Gaal, L. Forro, J. C. Bryan,
B. C. Chakoumakos, L. M. Woods, B. C. Sales, R. S. Fishman,
and V. Keppens, Phys. Rev. B.{\bf 63}, 195104 (2001).

\bibitem{furu} T. Furubayashi, T. Matsumoto, T. Hagino, and S. Nagata,
J. Phys. Soc. Jpn.{\bf 63}, 3333 (1994).

\bibitem{isobe} M. Isobe and Y. Ueda, J. Phys. Soc. Jpn. {\bf 71}, 1848
(2002). 

\bibitem{ogu} F. Ishii and T. Oguchi, J. Phys. Soc. Jpn. {\bf 69}, 526 (2000).

\bibitem{mielke} A. Mielke, J. Phys. A{\bf 24}, L73 (1991); {\bf 24},
3311 (1991).

\bibitem{ferro} K. Kusakabe and H. Aoki, Phys .Rev. Lett.{\bf 72}, 144
(1994).

\bibitem{graph} N. Biggs, {\it Algebraic Graph Theory}, 
(Cambridge Univ. Press, Cambridge, 1974).

\bibitem{fuji3} S. Fujimoto, Phys. Rev. Lett.{\bf 89}, 226402 (2002).

\bibitem{shan} R. Shankar, Rev. Mod. Phys. {\bf 66}, 129 (1994).

\bibitem{metz} C. J. Halboth and W. Metzner, 
Phys. Rev. B{\bf 61}, 7364 (2000).

\bibitem{kop} P. Kopietz and T. Busche, Phys. Rev. B{\bf 64}, 155101
(2001).

\bibitem{zan} D. Zanchi and H. J. Schulz, Phys. Rev. B{\bf 61}, 13609 (2000).

\bibitem{hone} C. Honerkamp, M. Salmhofer, N. Furukawa,
and T. M. Rice, Phys. Rev. B{\bf 63}, 035109 (2001).

\bibitem{frg} J. Polchinski, Nucl. Phys. B{\bf 231}, 269 (1984).

\bibitem{wet} C. Wetterich, Phys. Lett. B{\bf 301}, 90 (1993).

\bibitem{morris} T. M. Morris, Int. J. Mod. Phys. A{\bf 14},
2411 (1994). 

\bibitem{salm} M. Salmhofer, 
{\it Renormalization} (Springer, Berlin, 1998).


\bibitem{corr} Equation (4) in Ref. 43 should be corrected as
equation (\ref{rge3}) in this paper.
This correction does not affect the main results in Ref. 43 qualitatively,
but gives a quantative change in the estimation of $T_c$.

\bibitem{ex} B. I. Halperin and T. M. Rice, Solid State Physics {\bf 21},
115 (1968). 

\bibitem{al} K. Matsuno, T. Katsufuji, S. Mori, Y. Moritomo,
A. Machida, E. Nishibori, M. Takata, M. Sakata, N. Yamamoto,
and H. Takagi, J. Phys. Soc. Jpn. {\bf 70}, 1456 (2001).

\bibitem{harima} H. Harima, private communication.

\bibitem{tl} H. Sakai, M. Kato, K. Yoshimura, and K. Kosuge, 
J. Phys. Soc. Jpn. {\bf 71}, 422 (2002).

\bibitem{sin} D. J. Singh, P. Blaha, K. Schwarz, and J. O. Sofo, 
Phys. Rev. B{\bf 65}, 155109 (2002).

\end{references}
\end{document}